\documentstyle[12pt,epsfig]{article}
\begin{document}
\begin{center}
{\bf On One-Loop Gap Equations for the Magnetic Mass\\
in d=3 Gauge Theory}\\[.3in]
John M. Cornwall $^1$ \\[.2in]
{\it Department of Physics and Astronomy, University of California\\
405 S. Hilgard Ave., Los Angeles Ca 90095-1547}\\[.2in]
{\bf Abstract}
\end{center}
Recently several workers have attempted determinations of
the so-called magnetic mass of d=3 non-Abelian gauge theories through a
one-loop gap equation, using a free massive propagator as input.  Self-consistency is attained only on-shell, because the usual Feynman-graph construction is gauge-dependent off-shell.  We examine two previous studies of  the pinch technique proper self-energy, which is gauge-invariant at
{\em all} momenta, using a free propagator as input, and show that it leads to inconsistent and unphysical results.  In one case the residue of the pole has the wrong sign (necessarily implying the presence of a tachyonic pole); in the second case the residue is positive, but two orders of magnitude larger than the input residue, which shows that the residue is on the verge of becoming ghostlike.  This happens because of the infrared instability of d=3 gauge theory.  A possible alternative one-loop determination via the effective action also fails.  The lesson is that gap equations must be considered at least at
two-loop level.\\[1.5in]
UCLA/97/TEP/12 \mbox{} \hfill October 1997\\
\footnoterule
\noindent $^1$ Electronic address:  Cornwall@physics.ucla.edu\\
\newpage  
\section{Introduction}

Three-dimensional non-Abelian gauge theories are known \cite{l80,c82,chk85}
to generate a gauge-boson mass dynamically and non-perturbatively, although the mass vanishes to all orders of perturbation theory.  Such
theories constitute the only part of high-temperature gauge theories which
are perturbatively infrared-singular\footnote{In fact, they are much more than just 
singular by virtue of the vanishing of the perturbative mass; they are {\em infrared-unstable} \cite{c82} just as QCD is in d=4, and have tachyons in
perturbation theory.} and strongly-coupled.  In EW theory above the phase
transition temperature $T_c$, the Higgs VEV vanishes at lowest order and 
the magnetic EW gauge bosons are correspondingly massless perturbatively.
In this temperature regime perturbation theory \cite{pert} fails, although it is
perfectly applicable for all other Matsubara modes as well as for the $N=0$
electric sector (which gets a mass perturbatively).

In light of this problem, which stands in the way of our understanding of
early-universe processes, several authors have recently considered the
generation of a self-consistent magnetic mass (of $O(\alpha_WT)$ in the
EW sector), using one-loop gap equations \cite{an,bp1,bp2,jp1,jp2}.  The general thrust of
these works, which differ in detail, is to find some way of writing down
a gauge-invariant Lagrangian which generates mass, and then to construct a
one-loop gap equation in the spirit of similar fermionic equations used in
dynamical symmetry breakdown and in superconductivity.  While some works
\cite{bfhw} are not strictly gauge-invariant, the more recent approaches cited above (except for ref. \cite{an}, hereafter AN, which uses the pinch technique (PT) to maintain gauge invariance at all momenta)  maintain gauge invariance by working only on-shell\footnote{That the usual Feynman-graph self-energy gives gauge-invariant masses on-shell can be proved directly \cite{kkr}, and also follows from the PT \cite{c82,cp89} because this technique involves corrections to Feynman graphs that automatically vanish on-shell.}, that is, by
demanding self-consistency for the one-loop self-energy $\Pi (p^2)$ only
when $-p^2=m^2$ (we use Euclidean three-momenta throughout this paper).
This is to be contrasted to the use of gap equations for chiral symmetry
breakdown (for example), where one demands self-consistency for a continuum
of momenta.  

The question then arises:  Can self-consistency be addressed in the context
of a {\em one-loop} gap equation at all momenta, while maintaining gauge invariance?
Concerning gauge invariance, the answer is yes, with the use of the PT \cite{c82,cp89}.  This  technique was applied long ago \cite{chk85} (CHK hereafter) to the problem of finding a one-dressed-loop proper self-energy with the input matching the output at all momenta, and more recently by AN to finding the mass with a free massive propagator as input; in this latter work, no attempt was made to reach self-consistency except on the mass shell.  But concerning the specific application of the PT at one loop, CHK found that, without an {\it ad hoc} modification of the one-loop gap equation, intended to represent the contribution of two- and higher-loop graphs, no self-consistency could be found.  In the present work we extend these PT studies; for CHK, we do what all the other cited works do and use a free massive propagator as input, and for AN we study further the question of self-consistency.  The two cases are not identical:  Although the PT works exactly the same for both, the gauge-invariant mass term used in the self-energies differ somewhat, since there is no unique answer to constructing a gauge-invariant Lagrangian mass term.  In both cases we find serious difficulties, most easily characterized by the residue of the pole in the PT propagator.  For CHK we find that the residue is negative (ghost-like), which necessarily implies a second propagator pole which is tachyonic; for AN, the residue is positive, but 150 times as large as the (unit) input residue.  The great size of the AN residue is an indication that their propagator is just on the verge of showing a tachyon and a ghost-like residue.

It should be noted that in other recent works \cite{bp1,bp2,jp1,jp2} which study only the usual Feynman propagator, it makes no sense to ask for self-consistency of the residue, which is gauge-dependent.  But quite aside from this question of self-consistency, several of these authors had reason to be suspicious of one-loop gap equations; for example, Jackiw and Pi \cite{jp1} find a complex mass with their mass term, and point out the existence of unphysical (massless) states in the conventional Feynman propagator; these are absent in the PT
expressions (\cite{chk85,an}, equation (16) of the present paper).

As for the masses found in these various works, they also differ somewhat among themselves, for various reasons which we discuss below, but are in the range 0.124$\leq m/Ng^2\leq$ 0.192, for gauge group $SU(N)$, a range of about
$\pm$20\% around the mean of these two numbers.  This small range of variation would in itself be quite satisfactory, if it were not for the problems of lack of self-consistency we find.  

There are, in fact, several other reasons to be suspicious of one-loop gap equations, even those based on the PT.  In the first place, the range of masses given above is (for $SU(2)$) considerably smaller than lattice determinations \cite{kj,fhkjjm}, although here it is not clear that what is being found on the lattice is related to the propagator investigations.
Second, one knows, of course, that there is no identifiable small parameter which could be associated with a dressed-loop expansion (as remarked also by Jackiw and Pi \cite{jp2}).  Third, logarithmic divergences arise {\em in
perturbation theory} which can only be dealt with at two and more loops.
It was because of this that early work \cite{chk85} claimed only to be able
to give a lower limit on the magnetic mass, not a specific value. Finally, if the mass in the input propagator is
constant and not decreasing with momentum at large momentum, the whole program of mass generation does not work, as we illustrate with the CJT \cite{cjt} effective action.

The CJT approach yields a variational principle for the propagator, which could in principle be more successful than insisting on full self-consistency at all momenta because, as in quantum mechanics, variational wavefunctions (or self-energies, in our
case) can be rather more inaccurate than the variational energies coming from these wave functions.  As conventionally stated, this variational principle yields the Schwinger-Dyson equation for the Feynman propagator, but it can (at least at the loop level considered here) re-interpreted as a variational principle for the PT propagator.  This re-interpretation simply amounts to a re-arrangement of the usual CJT action in such a way that the value of the action at its minimum is not changed, but variation of the action with respect 
to a propagator yields the one-loop PT gap equation.  Such a rearrangement to get
the {\em massless} PT equations was considered some years ago \cite{c89}.  We will show that this approach
fails because of ultraviolet logarithmic divergences due to a constant magnetic mass.  In actuality the mass must vanish at large momentum.

While all of the above-mentioned problems are important, we have as yet barely touched upon the one really essential difficulty with one-loop gap equations in non-Abelian gauge theory: the infrared instability of such theories.  In three dimensions this instability is indicated by a wrong-sign contribution to the gauge-invariant PT propagator, which gives rise to a tachyonic pole even in perturbation theory \cite{c82,chk85}.  Generation of mass tends to remove this instability, but there is a critical value below which there are still tachyons, as CHK show for the fully self-consistent case.  Not surprisingly, the same problem emerges when a free massive propagator is used as input; only the details change.  For CHK's model of mass generation, the self-consistent one-loop mass is too small to get rid of the tachyon, while for AN's model it is just barely large enough to avoid tachyons.    Closely associated with the tachyonic pole of CHK is the equally-bad feature that the residue of the pole at $-p^2=m^2$ is negative and
unphysical.  The general lack of self-consistency arises because the input propagator contains terms likely to be characteristic of two-loop contributions (the mass term) while the output propagator has infrared-unstable one-loop contributions.

Our point is not the failure of the presently-formulated program of
finding a gauge-boson mass in d=3 gauge theory, but its failure at one-loop level.  In fact, we are hopeful that a cure is to be found in the (non-tachyonic, in general) contribution of 
two- and higher-loop contributions, which as mentioned above were inserted phenomenologically in CHK.

In the next section we give some details of how CHK is modified by using a free massive propagator as input; then after a brief comparison of self-consistent masses with other recent works \cite{an,bp1,bp2,jp1,jp2} we continue with a study of the properties of the AN model.

\section{Pinch-Technique Gap Equation}

The PT \cite{c82,cp89} begins from the S-matrix or other gauge-invariant
quantity and extracts from it gauge-invariant proper self-energies and
vertices---gauge-invariant at all momenta.  The standard example is shown in
Fig. 1, illustrating how the PT propagator is extracted from the on-shell
scattering of external particles (solid lines).  Figs. 1(a-d,f,g) are (some of)
the graphs.  Depending on the gauge chosen, the gauge propagators and vertices will have longitudinal momentum factors (e.g., $k_i$ in Fig. 1(d)) which
strike a vertex and generate an elementary Ward identity such as

\begin{equation}
k_i\gamma_i = S^{-1}(p)-S^{-1}(p-k).
\end{equation}
But $S^{-1}(p)=0$, and the other inverse propagator just cancels the line
carrying momentum $p-k$ in Fig. 1(d), leading to the pinch part Fig. 1(e),
which has the kinematics of a propagator insertion.  Similarly the box graph
and crossed-box graph may have pinch parts yielding Fig. 1(h), again contributing to a propagator structure.  The sum of the usual Feynman propagator parts and the pinch parts is gauge-invariant, as explicit calculations
\cite{c82,cp89} show.

To convert these purely perturbative concepts into a gap equation, we follow
the philosophy used previously \cite{c82,chk85}, which has its roots in the
original studies of dynamical symmetry breaking \cite{cn73,jj73}.  The underlying gauge theory is taken to be $SU(2)$, as in \cite{bp2}, but with no
other fields (B\"uchmuller and Philipsen include the Higgs field of EW theory,
which leads to some slight differences). We denote the PT propagator with its
gauge-invariant self-energy by $\hat{\Delta}$ to distinguish it from the usual
Feynman propagator $\Delta$.  Schematically, the one-loop gap equation
is:

\begin{equation}
\hat{\Delta}^{-1}=\Delta_0^{-1} -g^2\int \Gamma_0\hat{\Delta}\hat{ \Delta} \Gamma
+  \int \Gamma^{(4)}_0\hat{\Delta} +pinch\; terms 
\end{equation}
where $\Gamma_0$ is the usual bare vertex, $\Gamma$ is a full vertex which
carries information on mass generation through longitudinally-coupled 
massless poles (like Goldstone poles, but {\em not} associated with symmetry
breaking), and $\Gamma^{(4)}_0$ is the bare four-gluon vertex.  The pinch term of Fig. 1(e) is constructed with the full three-gluon vertex, and all pinch
terms have $\hat{\Delta}$ propagators.  

In principle the full vertex $\Gamma$ is to be determined from its own
Schwinger-Dyson equation, but we approximate it by a form which is expressed
solely in terms of the mass $m$ we are trying to determine, in such a way that
the PT Ward identities \cite{c82,cp89}

\begin{equation}
q_i(\Gamma -\Gamma_0)_{ijk}(q,p,-q-p)=\hat{\Pi}_{jk}(q+p)-\hat{\Pi}_{jk}(p).
\end{equation}
are satisfied.
In (3), $\hat{\Pi}_{jk}$ is the gauge-invariant proper self-energy associated
with the PT propagator:

\begin{equation}
\hat{\Delta}^{-1} = \Delta_0^{-1} - \hat{\Pi};
\end{equation}    
here $\hat{\Pi}$ is both gauge-invariant and conserved:

\begin{equation}
\hat{\Pi}_{jk}(p)=T_{jk}(p)\hat{\Pi}(p)
\end{equation}
with

\begin{equation} T_{jk}(p)=\delta_{jk}-p_jp_k/p^2.
\end{equation}

The general solution to (3) for any function $\hat{\Pi}$ is known \cite{ch86};
of course, it is indeterminate to the extent that any completely conserved function can be added to $\Gamma$. The models of CHK and AN differ only in this conserved term, and satisfy, as they must, the PT Ward identities.

Now we want to make contact with the recent work discussed above \cite{an,bp1,bp2,jp2}.
These authors assume that the propagator (for them, the usual Feynman propagator) can be approximated by that of a free particle of mass $m$:

\begin{equation} \Delta_{jk}(p)=\frac{T_{jk}(p)}{p^2+m^2} + gauge\;terms.
\end{equation}
We will use this form for the PT propagator $\hat{\Delta}$, which correspondingly requires the form of $\Gamma$ yielding the Ward identity
(3).  This has also been given in Ref. \cite{ch86} as a special case, and
we have

\begin{eqnarray}
\Gamma_{ijk}(q,p,-q-p)  =  (2p+q)_i\delta_{jk}-(2q+p)_j\delta_{ik}
+(q-p)_k\delta_{ij}+\\ -\frac{m^2}{2}[\frac{q_ip_j}{q^2p^2}(q-p)_k-\frac{p_j(p+q)_k}{p^2(p+q)^2}
(2p+q)_i+\frac{(p+q)_kq_i}{(p+q)^2q^2}(2q+p)_j] \nonumber
\end{eqnarray} 
which satisfies

\begin{equation}
q_i\Gamma_{ijk}(q,p,-q-p)=T_{jk}(p+q)(p^2+m^2)-T_{jk}(p)(p^2+m^2).
\end{equation}

While we intend that this form of $\Gamma$ is an approximation to the
full vertex of ordinary d=3 gauge theory, we note \cite{c74} that this is
the lowest-order vertex of the gauge theory to which has been added a
gauged non-linear sigma-model term, provided that the classical equations of motion of the sigma model are used to eliminate the sigma-model fields in terms of the gauge potentials.  This addition provides a gauge-invariant
mass at the classical level, although considered as an all-order field
theory it has problems of unitarity and renormalizability.  These problems are
of no concern to us, since we work only at one-loop level.  One can similarly
find a lowest-order four-gluon vertex, but this will not be needed here.
The action which generates the vertex of equation (8) is:

\begin{equation}
S=\int d^3x [-\frac{1}{2}TrG_{ij}^2-m^2Tr(U^{-1}D_iU)^2],
\end{equation}
where $D_i$ is the gauge covariant derivative.  If the classical sigma-model equations
\begin{equation}
[D_i,U^{-1}D_iU]=0
\end{equation}
 are used to eliminate $U$, and the (non-local) result expressed in terms of
the gauge potential is substituted back in the action (10), the lowest-order
vertex is just that of equation (8).\footnote{It therefore should be that the proper self-energy we calculate is the same as that of the gauged non-linear sigma model, where unlike equation (2) {\em all} vertices are those of
the non-linear sigma model to the appropriate order of $g^2$.  One can check that, because of the longitudinal structure of the massless poles in the vertex
(8) and the transverse structure of the gauge-independent part of the propagator, there is in fact no difference from what we present here; the only difference is a gauge artefact.}

It now remains to calculate, as in Refs. \cite{c82,chk85}, the gauge-invariant PT propagator for $SU(N)$ gauge theory.  After a lengthy calculation, we find

\begin{eqnarray} \hat{\Delta}^{-1}_{ij}(p)=T_{ij}(p)p^2-\frac{Ng^2}{(2\pi )^3}\int\frac{d^3k}{(k^2+m^2)((p+k)^2+m^2)}\times \\ \nonumber \times [T_{ij}(p)(4p^2+m^2)
+\frac{1}{2}(2k+p)_i(2k+p)_j] +\frac{Ng^2}{(2\pi )^3}\delta_{ij}\int
\frac{d^3k}{k^2+m^2}\\ \nonumber
+ gauge\; terms.
\end{eqnarray}
The gauge terms in (12) are purely kinematic, and come from the free propagator only; there is no gauge dependence in the self-energy.

With the aid of the integrals

\begin{eqnarray}
\int\frac{d^3k}{(2\pi )^3}[\frac{(2k+p)_i(2k+p)_j}{(k^2+m^2)((p+k)^2+m^2)}
-2\frac{\delta_{ij}}{k^2+m^2}]=\\ \nonumber
=-\frac{1}{2}T_{ij}(p)\int \frac{d^3k}{(2\pi )^3}[\frac{p^2+4m^2}{(k^2+m^2)((p+k)^2+m^2)} +\frac{2}{k^2+m^2}],
\end{eqnarray}

\begin{equation}\frac{1}{(2\pi )^3}\int \frac{d^3k}{( (k^2+m^2)((p+k)^2+m^2)}=
\frac{1}{4\pi p}\arctan (\frac{p}{2m}),
\end{equation}
and the dimensionally-regularized result

\begin{equation}
\frac{1}{(2\pi )^3}\int \frac{d^3k}{k^2+m^2}=-\frac{m}{4\pi},
\end{equation}
one comes to the final result for the conserved (modulo inessential gauge terms, which we drop) PT inverse propagator:

\begin{eqnarray}
\hat{\Delta}_{ij}^{-1}(p)  =  T_{ij}(p)[p^2-(\frac{Ng^2}{2})\frac{15p}{8\pi }
\arctan (\frac{p}{2m}) -(\frac{Ng^2}{2})(\frac{m}{4\pi})]\\
 \equiv T_{ij}(p)\hat{d}^{-1}(p^2) \nonumber
\end{eqnarray}
Note that in all these equations there are no threshholds corresponding to
physical propagation of massless particles, such as ghosts or the longitudinally-coupled massless poles in the vertex (8).  They all cancel when
the PT is used.  This is not the case for Jackiw and Pi's first formulation of
a gauge-invariant mass term \cite{jp1}.

By setting $-p^2=m^2$ one finds the self-consistent mass for $SU(2)$ (for
other $N$, multiply by $N/2$):

\begin{equation}
m=g^2[\frac{15}{16\pi}\ln 3 -\frac{1}{4\pi}] \simeq 0.248g^2
\end{equation}

\section{Comparison With Other Recent Work}

Now let us compare this mass value to others given recently.
Buchm\"uller and Philipsen (BP) \cite{bp2} calculated a magnetic mass for $SU(2)$
gauge theory with a Higgs field, and one can, it appears, remove the Higgs
field by letting its mass $M_H$ approach infinity, leaving only the massless
Goldstone fields (an equivalent calculation is done by Jackiw and Pi \cite{jp2}.)  At infinite Higgs mass, their result is:

\begin{equation}
m_{BP}=g^2[\frac{63}{64\pi}\ln 3-\frac{3}{16\pi}] \simeq 0.284g^2.
\end{equation}
This is larger than the value given in equation (17) by:

\begin{equation}
g^2[\frac{3}{64\pi}\ln 3+\frac{1}{16\pi}].
\end{equation}
Although the numerical difference is small (about 15\%), one may think that
the $M_H \rightarrow \infty$ limit should just reproduce the non-linear
sigma model result of (17).  Evidently this is not so.  The reasons are
somewhat subtle, having to do with lack of total decoupling in the
infinite-mass limit, and/or taking the $\xi \rightarrow \infty $ limit in the
$R_{\xi}$ gauge, thereby reaching the unitary gauge (as discussed by
Jackiw and Pi \cite{jp2}).  To note one (gauge-dependent!) distinction, 
observe that the massless limit of the BP self-energy in the Feynman gauge $\xi = 1$
(which can be found as the massless limit of equation (8) in \cite{jp2}) is not the same as the conventional massless
Feynman-gauge propagator.  The difference can be traced to the Goldstone-loop
integral, which for the massive theory in Feynman gauge contributes:
   
\begin{equation}
\frac{g^2}{4}\int \frac{d^3k}{(2\pi)^3}\frac{(2k+p)_i(2k+p)_j}
{(k^2+m^2)((k+p)^2+m^2)}.
\end{equation}
This does not vanish in the massless limit and in fact is responsible for the difference between the coefficients of the BP Feynman-gauge propagator and the
usual massless Feynman-gauge propagator.  Interestingly, it can readily be checked that
the mass contribution from (20) is just the first term in (19) (the rest is given by seagulls).  This is not a physical distinction, however, since in other $R_{\xi}$ gauges the mass in (20) becomes
$\xi m^2$ and this gauge-dependent term must be completely cancelled by
other contributions.

AN give a mass $m_{AN}$ which is 4/3 times the
BP mass (or about $0.38g^2$), and again different from (17) above.  In this case the discrepancy can
be traced to their form of a gauge-invariant mass term, differing somewhat from that of CHK.  

\section{Inconsistencies}

At this point we summarize both the above extension of CHK and AN's work in the following generic one-loop form of the PT propagator, using a free massive propagator as input:

\begin{equation}                                                 %21
\hat{d}^{-1}(p^2)=\frac{Ng^2}{4\pi}[-\alpha p \arctan (\frac{p}{2m})
-\beta m + \gamma \frac{m^2}{p}\arctan (\frac{p}{2m})].
\end{equation}
Note that there are no terms in (21) involving intermediate states other than two gluons of mass $m$ (such as $\arctan (2mp/m^2-p^2)$) coming from a massive and a massless state).  Only physical states enter the pinch-technique propagator.

For CHK the coefficients can be read off from (16), and it turns out that the values of $\alpha,\;\beta$ are the same as for AN:

\begin{equation}                   %22
\alpha = +15/4,\;\beta = 1/2.
\end{equation}
Clearly for CHK $\gamma$=0, while for AN we have
 \begin{equation}                                 %23
\gamma_{AN}= +3/2.
\end{equation}
The coefficient $\alpha$ is universal because it comes from the massless theory.  There is no deep reason for the other two coefficients to be the same or different, since the mass models differ, and in particular no significance should be attached to the vanishing of $\gamma$ for CHK.

Let us express the behavior of the {\em output} PT propagator near the mass-shell pole in terms of a residue $Z$:

\begin{equation}                             %24
\hat{d}(p^2) \simeq \frac{Z}{p^2+m^2},\;-p^2 \simeq m^2 .
\end{equation}
A small amount of algebra yields the following expressions for $m,\;Z$:

\begin{equation}                           %25
m=\frac{Ng^2}{4\pi}[(\frac{\alpha + \gamma}{2})\ln 3-\beta];
\end{equation}

\begin{equation}                          %26
Z^{-1}=\frac{g^2}{m}[\alpha (\frac{1}{4}\ln 3 -\frac{1}{3})-\beta
+\gamma (\frac{3}{4}\ln 3 - \frac{1}{3})].
\end{equation}
Of course, using the numerical values of $\alpha ,\; \beta ,\; \gamma$ in (25)
yields the previously-quoted mass values.  As for the residue, we find from
(26) that the CHK residue is {\em negative}, which is unphysical and which also
forces a tachyonic pole in the CHK propagator.  For AN, the residue has the
astoundingly large value of

\begin{equation}                       %27
Z_{AN} \simeq 150,
\end{equation}
compared to the input $Z$=1.

The upshot is that the PT expressions, both for CHK and for AN, for
the output propagator as a function of momentum are quite inconsistent with the
input free massless propagator.  In CHK the residue is ghostlike and there is a second tachyonic pole, as shown in Fig. 2, which plots the inverse PT propagator (labelled f on the figure) as a function of momentum.  In AN the residue is positive and there is no tachyon (as noted by AN), but the output residue is extremely large compared to that of the input.  This simply means that the AN model is very close to having a tachyonic pole.  

One might try a modified self-consistent scheme, finding values for $Z,\;m^2$ both of which are self-consistent, with an input propagator of the form (24).
This requires replacing the full vertex $\Gamma$ of the Schwinger-Dyson
equation (2) by $\Gamma /Z$ to maintain the Ward identity (3).  However, one can only find {\em complex } solutions for $Z$, which are certainly unphysical.

The essential reason for this behavior is that the coefficient $\alpha$, which is dominant numerically (more than twice as large as the other coefficients), has a positive sign and contributes negatively to the residue.  This positive sign is associated with infrared instability in the d=3 theory, which leads to a tachyonic pole even in perturbation theory ($m=0$).  The perturbative PT one-loop
propagator is:
\begin{equation}                      %28
\hat{d}^{-1}(p^2) = p^2-\frac{15Ng^2p}{32}.   
\end{equation}
The trouble is that the free propagator form (7) gives no hint of this
infrared instability.  But if it is used as input to the one-loop PT
integrals, which properly reflect the infrared problems of d=3 gauge theories, the instability returns.  It is, of course, true that mass generation is
essential to curing the infrared instability, but it need not be true that
this can be done at the one-loop gap equation level.  

We believe that it is difficult to overcome these problems with one-loop gap equations.  There is more than one reason.  In the
first place, to overcome the $O(g^2)$ terms in the CHK PT propagator (16), which are both negative, one needs some positive loop contributions.  These are readily
found at two loops, but not at one loop.  In the second place, it could be
(and was \cite{c82,chk85}) argued that if a free massless propagator is not
a good approximation, perhaps one should use a propagator for the gap equation
which is massive, but which reflects the incipient infrared instability of perturbation
theory.  In Refs. \cite{c82,chk85} this was attempted with the aid of the
so-called gauge technique, which represents the propagator as a spectral
integral over masses and solves the Ward identities in a corresponding integral
fashion.  The result was a non-linear one-loop gap equation in which one could try to solve for the PT propagator self-consistently at all momenta, rather than simply checking whether a predefined input like the free propagator (7) led to a self-consistent mass.  Sensible (tachyon-free) solutions could only be found by the {\it ad hoc} addition
of a positive term to the inverse PT propagator equation, interpreted as
$m^2(0)$, the squared ``mass" at zero momentum (this is a gauge-invariant
concept for the PT, but not \cite{kkr} for the usual Feynman propagator.
This added term is necessarily $O(g^4)$.  To get rid of the tachyon, $m(0)$
had to be bigger than a lower limit (found numerically) of

\begin{equation} m(0) \geq 1.96\frac{15Ng^2}{32\pi} \simeq 0.58 (\frac{Ng^2}
{2}),
\end{equation} 
substantially bigger than the one-loop values discussed so far.
This {\it ad hoc} term is clearly something like a two-loop contribution from
a seagull graph, but here one encounters another difficulty, if one hopes to
interpret this as coming from modifying the input to the one-loop gap equation
by changing the former free massive propagator (7) input to something with the
correct one-loop behavior at large momentum.  For example \cite{c82,chk85} if
we just use the perturbative propagator (28) in the seagull graph we encounter

\begin{equation}                      %29
g^2\int d^3k\frac{1}{k^2-15Ng^2k/32}\rightarrow  g^2\int d^3k[\frac{1}{k^2}
+\frac{15Ng^2}{32k^3}+ \cdots ].
\end{equation}            
Of course, the $1/k^2$ term is removed by dimensional regularization, but
the next term cannot be so removed, and unfortunately it is ultra-violet
divergent.  This divergence, we know, is cancelled by some other two-loop
graph in perturbation theory (and {\it a fortiori} in a correct gap
equation), but so far no one has investigated any kind of two-loop gap equation.
In consequence, it makes no sense to try a one-loop gap equation of the PT 
type without knowing how to regulate the $O(g^4)$ seagull divergences.  

\section{Another Hope Dashed}

A good part of the problem with one-loop gap equations of the PT type stems
from trying to match an input with an output locally in momentum space.  
Variational techniques, such as those of the CJT effective action \cite{cjt},
offer a hope of overcoming such a difficulty since these are global and not
local.  Just as in ordinary quantum mechanics, one may get decent results for
a variational estimate of a global quantity even thought the local input
(trial wavefunction) is not very good.  In this case the global quantity
is the vacuum energy.  We know \cite{c94} the exact form of the effective
action in its dependence on the zero-momentum component of the condensate
variable $Tr G_{ij}^2$, but there is an undetermined constant, proportional
to $g^6$, which is the VEV of this variable.  A successful application of the
variational principle would determine this constant.    

Following earlier work \cite{c89} which set up a CJT variational principle
for the {\em massless} propagator of equation (28), it is easy to find a
variational principle which yields the one-loop gap equations of the PT.  We give the details only for the CHK propagator.  The
difference between the usual CJT variational principle, which yields the  
Feynman-propagator Schwinger-Dyson equation, and the PT principle given here,
is merely a rearrangement of terms in the CJT action, just as the PT itself
is a rearrangement of terms in the S-matrix.  The value of the action at the
minimum remains unchanged.  To find a one-loop gap equation we need a two-loop
effective action.  

It turns out that the needed form of the CJT action $\Gamma$ (where $Z=\exp \Gamma$)is:

\begin{eqnarray}
\frac{(2\pi)^3\Gamma}{(N^2-1)V} & = & \int d^3k\{ \ln \hat{d}d_0^{-1}
-\hat{d}d_0^{-1}+1\}  \\ \nonumber
& & +\frac{5Ng^2}{4(2\pi )^3}\int \stackrel{3}{\prod} [d^3k_i\hat{d}(k_i)]\sum k_i^2
\delta (\sum k_i)\\ \nonumber
& & -\frac{5Ng^2m^2}{2(2\pi )^3}\int \stackrel{3}{\prod} [d^3k_i\hat{d}(k_i)]\delta
(\sum k_i)  +\frac{4Ng^2}{(2\pi )^3}[\int d^3k \hat{d}(k)]^2.
\end{eqnarray}
Here $\hat{d}$ is the coefficient of the transverse projector in the
PT propagator, and $V$ is the volume of space.  Variation of $\hat{d}$ in this effective action leads to
equation (12) (with the projector factored out) {\em provided} that after
variation $\hat{d}$ is replaced by the free propagator (7).  We wish to
evaluate the effective action as a function of $m^2,g^2$ and look for minima
when $m^2$ is varied. 

Unfortunately, this project is doomed to failure because of logarithmic 
ultraviolet divergences, of the form $m^2g^2\ln (\Lambda^2/m^2)$.  This is,
of course, just the sort of divergence one expects from evaluating a two-loop
integral, as we have illustrated above.  But in this case the divergence is
not associated with perturbative effects, because it is multiplied by $m^2$.

This behavior is closely related to a point made long ago in the days of (d=4) dynamical symmetry
breakdown \cite{cn73,jj73}:  {\em Dynamical generation of a mass which is not
allowed by a symmetry principle (in our case, gauge invariance) only works
if the mass is a function of momentum, vanishing at large momentum.}  Furthermore, only if the mass vanishes in just the right way will the effective action be finite, as a cancellation of terms in the effective action (an illustration is also given in Ref. \cite{cjt}).  If the magnetic mass does vanish at large momentum the logarithmic divergences we have encountered will go away.  
The need for the vanishing of the d=3  magnetic mass at large momentum has been discussed before \cite{c82,chk85,ch86}, but these papers came to no definite conclusion.
More recently, Lavelle \cite{l91} has shown, using the PT for the gauge propagator in the presence of a condensate, that in both d=3 and d=4 the
dynamical gauge-boson mass vanishes at large $p$ like $\langle Tr G_{ij}^2
\rangle /p^2$.  Evidently, this would cure the divergence in the CJT two-loop action.  However, we are not aware of any one-loop gap equation treatment which would insure this vanishing.

\section{Conclusions}

The prospects for any reasonable estimate of the magnetic mass by one-loop
gap equations seem dim.  When one attempts to extend on-shell self-consistency to all momenta with the PT, the output PT propagator looks very unlike the
input free massive propagator, and shows either a tachyon (with wrong-sign residue for the ``physical" pole) or an incipient tachyon, with a huge physical-pole residue.  These effects stem from the
very infrared instability that mass generation is trying to cure.  

If one then uses as an input a PT propagator with the correct $O(g^2)$
behavior at large momentum, then the seagull graph develops an ultraviolet
logarithmic singularity which is purely perturbative.  This sort of
perturbative singularity must ultimately cancel, but because it is $O(g^4)$
this can only be done at two-loop order.

A similar singularity shows up in a variational effective action, preventing
its use.  The cause this time is non-perturbative, and is traced to the failure
of the variational mass to vanish at large momentum, as is required for the
success of the mass-generation program.

A two-loop gap equation offers interesting prospects for curing these diseases
by adding a sufficiently large positive term to the inverse propagator (as modelled phenomenologically in Ref. \cite{chk85}).  It
is essential to study the two-loop problem not because it might yield increased
accuracy, but because the one-loop problem seems to be fatally sick.

Another possibility is an approach in the spirit of that of Karabali and Nair
\cite{kn}, which is not tied to loop expansions.

\section{Acknowledgements}

I thank Roman Jackiw, W. Buchm\"uller, V. P. Nair, O. Philipsen, and So-Young Pi for interesting correspondence concerning
their ideas.  This work was supported in part by the National Science Foundation under grant PHY 9531023.

\newpage
{\Large \bf Figure Captions}\\

\bigskip

Fig 1.  Some of the on-shell graphs for scattering  by gauge bosons at one-loop level, and the associated pinch contributions (e,h).

Fig. 2.  A plot of the inverse propagators ($f(p^2)$, in units of $m^2$) versus $p^2/m^2$ for the
input (equation (7); straight line) and output (equation (16)). 
\newpage
\epsfig{file=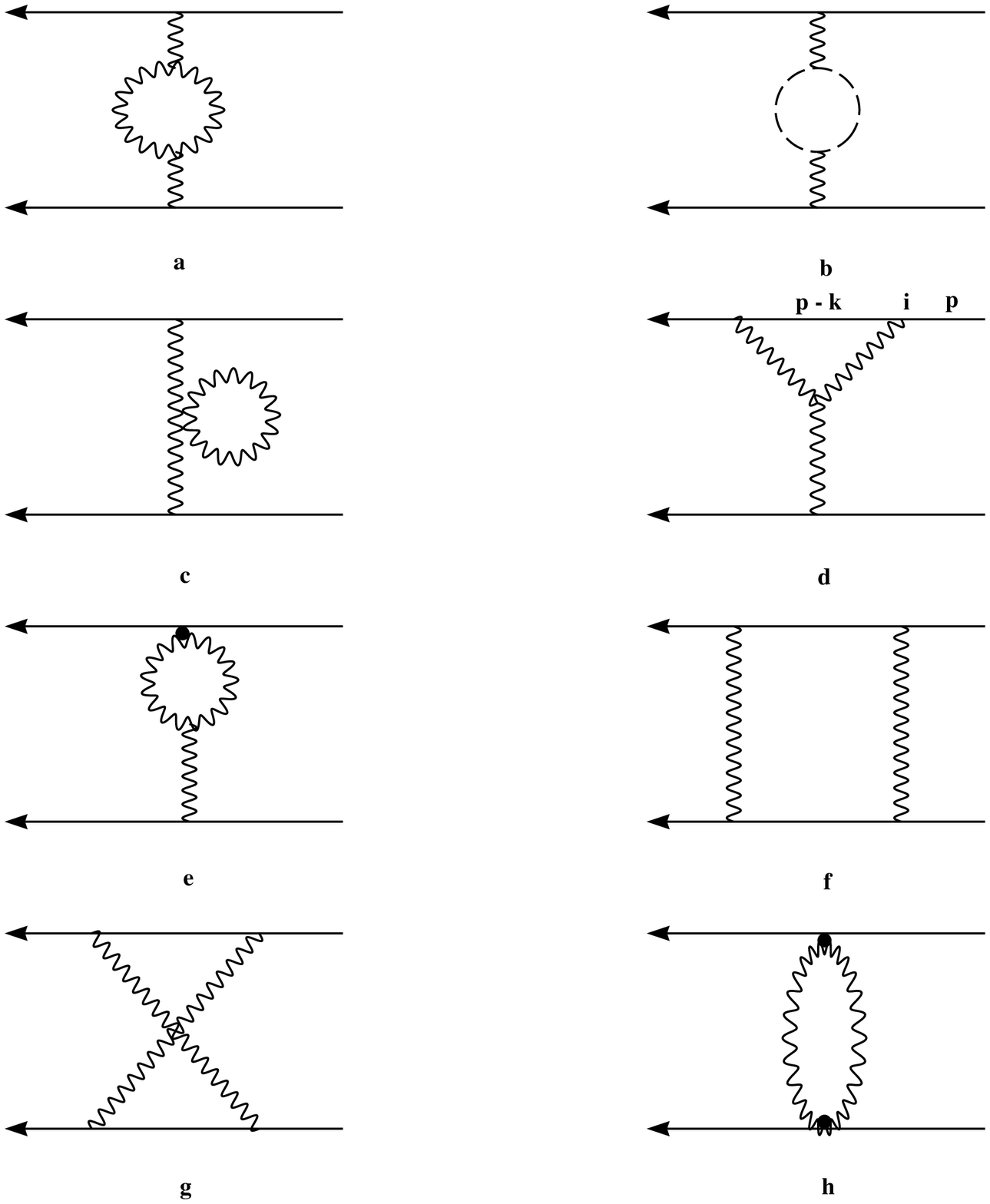,height=6in,clip=}
\newpage
\epsfig{file=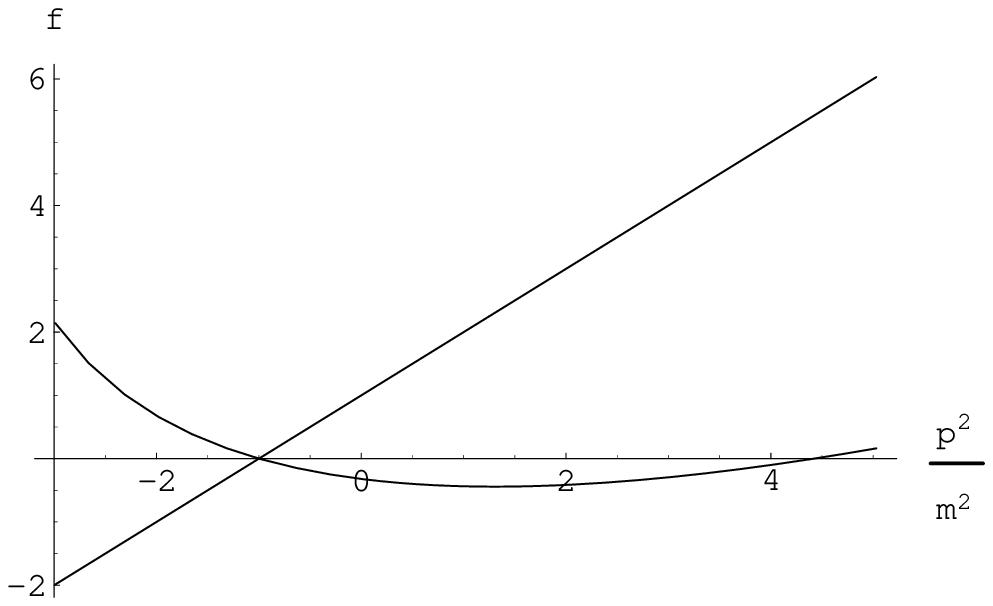,clip=}
\end{document}